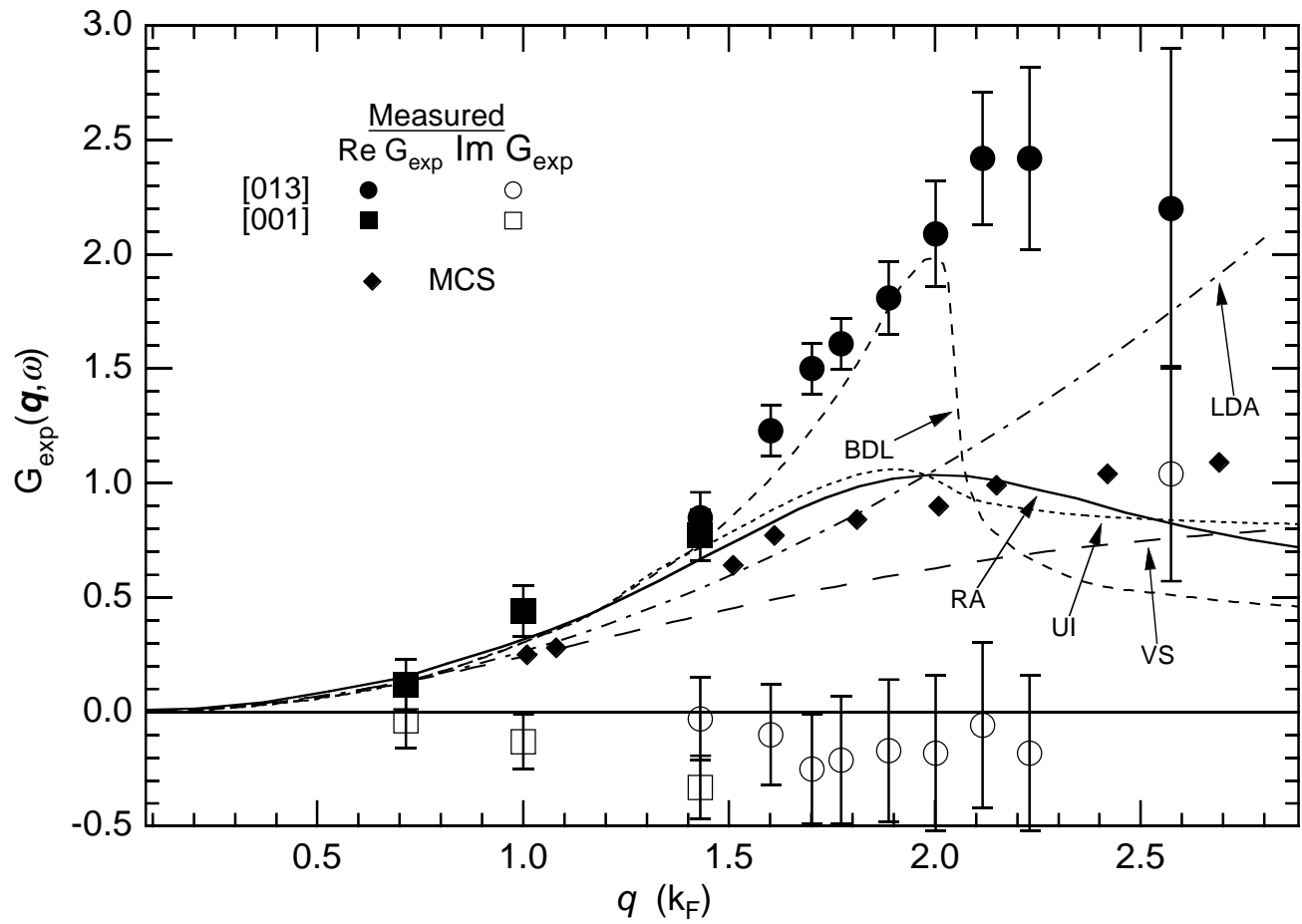

Fig. 1

# Electronic Excitations and Correlation Effects in Metals

Adolfo G. Eguiluz[*] and Wolf-Dieter Schöne

*Department of Physics and Astronomy, The University of Tennessee, Knoxville, TN 37996-1200*

*and Solid State Division, Oak Ridge National Laboratory, Oak Ridge, TN 37831-6030*

[*]  Corresponding author; e-mail: eguiluz@utk.edu

# Abstract

Theoretical descriptions of the spectrum of electronic excitations in real metals have not yet reached a fully predictive, "first-principles" stage. In this paper we begin by presenting brief highlights of recent progress made in the evaluation of dynamical electronic response in metals. A comparison between calculated and measured spectra —we use the loss spectra of Al and Cs as test cases— leads us to the conclusion that, even in "weakly-correlated" metals, correlation effects beyond mean-field theory play an important role. Furthermore, the effects of the underlying band structure turn out to be significant. Calculations which incorporate the effects of *both* dynamical correlations and band structure from first principles are not yet available. As a first step towards such goal, we outline a numerical algorithm for the self-consistent solution of the Dyson equation for the one-particle Green's function. The self-energy is evaluated within the shielded-interaction approximation of Baym and Kadanoff. Our method, which is fully conserving, is a finite-temperature scheme which determines the Green's function and the self-energy at the Matsubara frequencies on the imaginary axis. The analytical continuation to real frequencies is performed via Padé approximants. We present results for the homogeneous electron gas which exemplify the importance of many-body self-consistency.

Most properties of metals are strongly influenced by the electron-electron interactions [1]. For example, these interactions are responsible for the existence of collective excitations, such as plasmons and spin waves; without exchange and correlation there would be no metallic cohesion, or magnetism in the 3*d* transition metals, etc.

The theoretical treatment of correlation has traditionally been restricted to "simple models" which, by design, isolate some of the features of the problem which are deemed to be important. Now, approximations at two different levels are actually built into the models. First, a compromise is made in the description of the underlying band structure. In the jellium model, the band structure is simply ignored altogether —the electrons propagate in plane-wave states. This model has played a time-honored role in the study of correlation in simple metals. In the opposite end we have the Hubbard model [2], which corresponds to a tight-binding description of the band structure, in which, e.g., the hybridization of *sp* and *d* orbitals at the Fermi surface is neglected. This model was originally proposed for the study of transition-metal magnetism, and has been much-invoked in recent years for the study of highly-correlated electrons (e.g., in the cuprate high-temperature superconductors).

Second, a "model" or an approximation is introduced for the actual description of dynamical correlations. Thus, in the case of electronic excitations in jellium, the Coulomb interaction is often treated in a mean-field sense, such as the random-phase approximation (RPA) [1]. Short-range correlations are usually added on in simplified ways [3]. In the case of the Hubbard model, recent progress has been made in the treatment of correlation processes beyond mean-field theory. These include self-consistent diagrammatic approaches [4-6] — which, in fact, provide motivation for our work described below— and non-perturbative treatments of the Coulomb interaction via Quantum Monte Carlo methods [6], and exact diagonalization of the Hamiltonian for small clusters [7]. However, the long-range aspects of the Coulomb interaction are typically ignored in these schemes, which allow the electrons to interact only when they encounter each other at a given atomic site with opposite spin projections.

Of course, if the compromise contained in the above models at the level of the band structure is considered to be unreasonable for the problem at hand, one still has available the powerful method of density functional theory (DFT), which has the great appeal that the electronic structure is dealt with in a very realistic way [8,9]. However, in typical implementations of DFT the correlation problem is basically "taken for granted," in the sense that one assumes the validity of the local-density approximation (LDA), or gradient corrections thereof. At the level of the LDA we rely on an available approximate solution of the correlation problem for electrons in jellium. While the DFT method has proved extremely successful over the last two decades [10-12], its realm is basically confined to ground state observables which are obtainable from a knowledge of the total energy of the system.

In the first part of this article we briefly discuss some key results of recent work on the spectrum of elementary excitations of *sp*-bonded metals such as Al and Cs [13,14]. The theoretical work goes beyond the simple-model stage in the sense that the electrons are allowed to propagate in the "actual" band structure of the metal [13-21] (the band structure is, of course, that obtained in the LDA). However, the treatment of correlation is still at the mean field level, and there is no attempt at self-consistency. Nonetheless, the fact that the band structure is dealt with realistically allows us to establish a useful interplay with the results of modern spectroscopic measurements on these systems [22,23]. The end result is that we actually learn new physics and pose new questions. For example, it becomes clear that the effects of the band structure can be significant, even for these otherwise jellium-like systems [13-18,20]. In particular, this interplay has led to the experimental determination of the so-called many-body local field factor of Al [23,24]. This quantity condenses the effect of correlations beyond the RPA; its measured value [23] differs from theoretical predictions for wave vectors $q \sim 2k_F$, where $k_F$ is the Fermi wave vector.

Thus, even for weakly-correlated, "jellium-like" metals, a complete treatment of the excitation spectrum must include not only the effects of the band structure, but it must also incorporate the effects of correlation on the same footing. Little is known quantitatively about this joint problem beyond the description contained in the LDA.

In the case of semiconductors, a rather large volume of work has been devoted to a description of the impact of dynamical correlations on the fundamental band gap and quasiparticle energy-shifts of the one-electron band structure [25-29]. As is well known, the energy difference between the lowest unoccupied and the highest occupied Kohn Sham eigenvalues deviates from the experimental value of the band gap by 50-100%. This problem has been addressed in recent years at the level the so-called *GW* approximation [26-29], which yields results in apparent *quantitative* agreement with experiment. (We note that some of these calculations contain one or more ad hoc approximations —such as the neglect of the damping processes contained in the imaginary part of the self-energy, and the use of a plasmon-pole approximation; furthermore, the many-body requirement that the propagators must be dressed fully self-consistently with the self-energy has typically been neglected.) Recent calculations [30] have included the actual frequency dependence of the polarizability in the evaluation of the *GW* self-energy, with equal degree of success —as measured by agreement with the experimental band gap. Some work along similar lines has been performed for metals. For example, the self-energy effects in the occupied bandwidth of simple metals [31] and the exchange splitting of the magnetic bands of Ni [32] have been calculated; the agreement with experiment in this case is not as good as it is in the case of semiconductors.

With the above material as background and motivation, we move on to discuss ongoing work which constitutes a first step towards a realistic description of correlation in metals. We report results of a fully conserving solution of the Dyson equation for the one-particle Green's function within the shielded-interaction approximation [33]. That the solution is conserving means that it obeys important microscopic conservation laws [33,34]; technically, the propagators are dressed self-consistently with the self-energy. We find that many-body self-consistency is important. We illustrate this conclusion with numerical results for the spectral function, the quasiparticle weight at the Fermi surface, and the density of states. Of course, the screened-interaction approximation ignores additional correlation effects (e.g., renormalized vertices). More general self-energy functionals will be addressed elsewhere.

In this preliminary account of our method for the treatment of correlation in metals we confine the discussion to the homogeneous electron gas, or jellium model. However, our finite-temperature many-body techniques are applicable for realistic band structures. In fact, the calculations reported in this article correspond to an "empty-lattice" treatment of the band structure, since we sample wave vector space over a discrete three-dimensional mesh. Calculations for actual metals are in progress.

Accurate results for the Green's function, polarizability, self-energy, etc., are obtained by the implementation of a procedure which efficiently minimizes the impact of the frequency cutoff in the evaluation of the Green's function for imaginary times. The polarizability for imaginary times is then obtained as a product of two Green's functions (in wave vector space we perform a convolution), and is subsequently fast-Fourier transformed to the Matsubara frequencies. The analytical continuation of the self-energy to real frequencies is performed via Padé approximants [35].

## Dynamical Response in Real Metals: A Brief "Status Report"

The latest generation of calculations of dynamical electronic response in metals has reached a new level of sophistication [13-21]. Indeed, it is now possible to account for the effects of the one-electron band structure (as produced by the LDA) with great detail. In conjunction with significant developments on the experimental front (greatly enhanced energy and momentum resolutions have become available; improved sample-preparation techniques have made possible the realization of experiments yielding much "cleaner" data) the new theoretical algorithms allow us to delve into the physics of the excitations to a degree which was not possible until recently. As a result of this feedback between theory and experiment, new physical mechanisms are coming to the fore, as we sketch briefly next.

The comparison between measured and calculated cross sections for the inelastic scattering of, e.g., electrons and x-rays, can be conveniently formulated in terms of the dynamical structure factor [1]

$$S(\vec{q};\omega) = -2\hbar\Omega_N \, \text{Im}\, \chi_{\vec{G}=\vec{0},\vec{G}'=\vec{0}}(\vec{q};\omega) \,, \tag{1}$$

where $\chi$ is the dynamical density-response function. In the ab initio work performed so far [13-21], which is basically "RPA-like," the response function is given by [24]

$$\chi = P^{(0)}\left\{1 - v(1-G)P^{(0)}\right\}^{-1}, \tag{2}$$

where $v$ is the bare Coulomb interaction. In Eq. (2) we have used symbolic notation; in the present discussion, aimed at simple metals, it should be thought of as a matrix equation in the Fourier representation which arises naturally in the scattering problem implicit in Eq. (1). In Eq. (2) we have introduced the many-body local-field factor $G(\vec{q};\omega)$ [24] which (following the early work of Hubbard) is defined such that it accounts for all the effects of exchange and short-range correlations [36]. The Fourier transform of the non-interacting polarizability $P^{(0)}$ entering Eq. (2) is given by the well-known formula

$$P^{(0)}_{\vec{G},\vec{G}'}(\vec{q};\omega) = \frac{1}{\Omega_N} \sum_{\vec{k}}^{BZ} \sum_{n,n'} \frac{f_{\vec{k},n} - f_{\vec{k}+\vec{q},n'}}{E_{\vec{k},n} - E_{\vec{k}+\vec{q},n'} + \hbar(\omega + i\eta)} \langle \vec{k},n | e^{-i(\vec{q}+\vec{G})\cdot\hat{\vec{x}}} | \vec{k}+\vec{q},n' \rangle$$

$$\text{x} \quad \langle \vec{k}+\vec{q},n' | e^{i(\vec{q}+\vec{G}')\cdot\hat{\vec{x}}} | \vec{k},n \rangle \,, \tag{3}$$

about which we will have more to say later on. (In Eqs. (1) and (3) $\Omega_N$ denotes the volume of the periodically-repeated "cluster" on whose sides we apply Born–von Karman periodic boundary conditions). Equations (1)-(3) form the basis of several recent studies. Here we shall only touch on two test cases, which have proved quite instructive.

*The many-body local-field factor of Al*

An enormous number of calculations of the many-body local field factor have been reported in the literature over the years. Most calculations refer to the static limit, in which the exchange-correlation hole —whose physics is accounted for by the presence of $G(\vec{q};\omega)$ in the response function given by Eq. (2)— is assumed to adjust instantaneously as its "parent" electron propagates through the system.

The availability of ab initio results for the non-interacting polarizability $P^{(0)}_{\vec{G},\vec{G}'}(\vec{q};\omega)$ given by Eq. (3), computed for the LDA band structure, suggests that a *measurement* of the loss spectrum (which is basically given by $\operatorname{Im}\chi$, according to Eq. (1)), followed by an "inversion" of the data, would lead to an experimental determination of $G(\vec{q};\omega)$ via Eq. (2) [13]. Because of error-propagation in the data inversion, both the experimental measurements and the calculated values of the polarizability must of high quality. This procedure was implemented recently by Larson et al. [23], who performed measurements of $S(\vec{q};\omega)$ for Al over a large wave vector domain, with particular emphasis on the crucial regime $q \approx 2k_F$.

The experimental result for $G(\vec{q};\omega)$ is shown in Fig. 1. For illustration purposes, we compare the measured local-field factor with a small subset of the available theoretical results for this quantity (its zero-frequency limit) calculated for jellium with the average density of Al, $r_S = 2.07$; these are, the many-body results of Vashishta and Singwi (VS) [37], Utsumi and Ichimaru (UI) [38], Richardson and Ashcroft (RA) [39], and Brosens, Devreese and Lemmens (BDL) [40], as well as the local-density approximation (LDA) [16], and the recent Quantum Monte Carlo simulations of Moroni, Ceperley and Senatore (MCS) [41]. (The experimentally-determined $G(\vec{q};\omega)$ turns out to be predominantly real, and largely frequency-independent, over a rather wide energy interval, $10 \leq \hbar\omega \leq 40 \text{eV}$; thus, the comparison with static local-field factors is reasonable, even if preliminary.)

Clearly, the theoretical values of the many-body local-field factor are in good agreement with experiment up to $q \sim 1.5 k_F$. The data of Larson et al. [23] are also consistent with the plasmon dispersion relation calculated by Quong and Eguiluz [15] in a time-dependent extension of local density functional theory (TDLDA). (That is, the local-field factor implicitly built into the dispersion relation of the Al plasmon obtained in Ref. [15] agrees with the measurements of Larson et al. [23] for the wave vectors for which the response of the electrons is predominantly collective.) However, for the larger wave vectors for which the electronic response is incoherent, in particular, for $q \to 2k_F$, the experimental $G(\vec{q};\omega)$ differs significantly from —*it becomes much larger than*— the theoretical predictions.

[Although, as seen in Fig. 1, the result of BDL [40] agrees closely with the measured $G(\vec{q};\omega)$ up to $q \sim 2k_F$, the significance of this agreement (which does not subsist beyond $2k_F$) is not obvious, since the calculations of BDL did not account for the screening of the exchange ladders. Furthermore, a dynamical $G(\vec{q};\omega)$ obtained by the same group [40] differs markedly from experiment.]

The above finding is indicative of the existence of significant many-body correlations in this prototype of "simple and weakly-correlated metal" behavior —and it highlights the fact that theory still does not have predictive power in the treatment of dynamical correlations in metals, particularly at large wave vectors. Additional work along similar lines [42] further reinforces the message that correlation must be tackled on the same footing with a realistic description of the underlying band structure, i.e., a simple model like jellium does not suffice. And neither does the simple LDA description of correlation, even if its adoption allows the use of realistic band structures.

## *The plasmon dispersion relation in Cs*

The dispersion relation of the plasmon in the heavy alkali metal Cs, measured by vom Felde et al. [22] via high-resolution electron energy loss spectroscopy, is in *qualitative* disagreement with textbook physics [1]. The RPA for the density-response function, which corresponds to Eq. (2) with $G = 0$, is expected to be accurate in the small-wave vector limit, in which the bubble diagrams dominate the polarizability. This mean-field approximation yields a quadratic dispersion relation for small $q$'s (of course, with positive curvature), in qualitative accord with experiments performed over the years for many *sp*-bonded elements —which, to a good extent, accounts for the popularity and usefulness of the RPA for these systems. However, in the case of Cs, the dispersion relation of the plasmon turns out to have a *negative slope* for small wave vectors [22]. Moreover, for large wave vectors the dispersion relation is quite flat, in sharp contrast with the strong dispersion predicted by available theories of correlation for electrons in jellium with the bulk density of Cs, $r_s = 5.6$.

The original interpretation of the experiment of vom Felde et al. [22] was that it provided a signature of the presence of strong electron-electron correlations. Since the restoring force for the plasmon is related to the compressibility of the electron gas, the negative dispersion appeared to raise the issue of the stability of the system —a possible scenario would be a tendency towards Wigner-crystal formation. Of course, since Cs has the lowest valence-electron density of all elemental metals, the question of the importance of correlation suggests itself a priori.

As we now know [14], short range correlations do play a role in the present problem, but the same is different from the initial conjecture [22]. Indeed, the negative slope of the plasmon dispersion in Cs has been shown to be a band-structure effect [18]. More specifically, this effect has been traced to the contribution to the polarizability $P^{(0)}_{\vec{G},\vec{G}'}(\vec{q};\omega)$ from one-electron transitions to a complex of final states which are nearly-degenerate with the plasmon energy $(\sim 3\text{eV})$ [14]. Since a pedagogical discussion of this effect has been given elsewhere [36], and in keeping with the theme of this article, we move on to sketch the way in which the effects of correlation are contained in the experimental data.

Figure 2 shows calculated dispersion curves for the Cs plasmon. It is significant that the polarizability given by Eq. (3) was obtained from an LDA band structure in which the 5$p$ orbitals were treated as valence states. (To this end an ab initio pseudopotential was constructed for the ionic configuration 5$p$6, 6$s$0.7 [14].) The left panel of Fig. 2 shows the plasmon dispersion curve obtained from a scalar version of Eq. (2), in which we only keep the $\vec{G} = \vec{G}' = 0$ element of $P^{(0)}$. Such solution ignores the "crystal local fields," i.e., the contribution to the screening from density fluctuations of wavelengths comparable with the lattice constant. For small $q$'s the effect of the many-body local field, or "vertex correction" $f_{xc} = -vG$, is negligible. This is as expected, since, as recalled above, in this limit the bubble diagrams dominate the response. The agreement with the experimental data shown in Fig. 2 is excellent —in fact, it is "too good," as we indicate momentarily.

The very inclusion in the polarizability of spatially-localized orbitals —and these semicore states do contribute to the calculated plasmon dispersion for small wave vectors— immediately prompts the question of whether the crystal local fields can really be ignored. They cannot. Dispersion curves computed in the presence of the crystal local fields are shown on the right panel of Fig. 2. Three sets of calculations are actually represented (as was the case on the left panel), corresponding to different approximations for the treatment of the electron-electron correlations. These are, respectively: (i) RPA, for which $f_{xc} = 0$, (ii) TDLDA, for which $f_{xc} = \int d^3 x \, e^{-i\vec{q}\cdot\vec{x}} \, dV_{xc}(\vec{x})/dn(\vec{x})$, where $V_{xc}(n)$ is the exchange-correlation potential obtained in the LDA ground-state calculation, and (iii) Vashishta-Singwi (VS), who obtained an approximate vertex function from a decoupling of the equation of motion for the electron-hole pair density-fluctuation operator [37] (this vertex was already considered in Fig. 1). It is quickly apparent that Fig. 2 (right panel) presents us with a surprise: *the calculated dispersion relation contains a sizable correlation effect for small wave vectors*.

The explanation of this result is as follows. The loss spectrum $S(\vec{q};\omega)$ (whose main peak defines the energy position of the plasmon for a give wave vector) is obtained upon inverting the matrix $\left(1 - v(1-G)P^{(0)}\right)^{-1}$. The inversion of this matrix gives rise to a "feedback" between *large* wave vector arguments $\vec{q}+\vec{G}$ in $P^{(0)}$ and the *small* wave vector $\vec{q}$ of the plasmon (we recall that $P^{(0)}{}_{\vec{G},\vec{G}'}(\vec{q};\omega) \equiv P^{(0)}(\vec{q}+\vec{G},\vec{q}+\vec{G}';\omega)$). At the level of the RPA this crystal local-field effect shifts the energy of the plasmon *upwards*. This is a purely kinetic-energy (or electron-gas pressure) effect. By contrast, the introduction of a vertex correction $f_{xc} = -vG$ in Eq. (2) induces a *downward* shift of $\omega_p$; this is a physical consequence of the weakening of the screening due to the presence of an exchange-correlation hole about each screening electron. Note that the reason that this mechanism becomes operative for small $q$'s is the feedback induced by the crystal local fields, which ultimately originates in the contribution from the localized semicore charge to the dynamical polarizability.

We emphasize that, while we are in the presence of a correlation effect for small wave vectors, the same is quite different from the initial assumptions about the origin of the negative plasmon dispersion in Cs [22]. This example illustrates again the importance of ab initio calculations of dynamical response. Without a realistic description of the effects of the band structure, the interpretation of the experimental data becomes clouded by built-in assumptions (or "prejudices") stemming from simple model descriptions (e.g., electrons in jellium).

Note also that the above treatment of correlation, via "off the shelf" many-body local field factors, is far less than "fundamental." The merit of the above procedure is simply that it illustrates the fact that both in the present problem, and in the case of the large-wave vector response of Al, correlation effects play a quantitatively important role. In both problems, an accurate, first-principles theory of dynamical correlations for electrons propagating in the actual band structure of the system has yet to be developed.

## Back to the Beginning: Self-Consistent Solution of the Dyson Equation

A rigorous formulation of the many-body problem of interacting electrons starts out from the Dyson equation for the one-particle Green's function [24,33], which we write down as

$$G^{-1}(\vec{q};i\omega_n) = G_0^{-1}(\vec{q};i\omega_n) - \Sigma_{xc}(\vec{q};i\omega_n) , \qquad (4)$$

where we have Fourier-transformed the space- and time- dependence of all quantities. We use the finite-temperature Matsubara formalism [24], in which the Green's function for imaginary-times is Fourier-analyzed according to

$$G(\vec{q};i\omega_n) = \int_0^{\beta\hbar} d\tau \, e^{i\omega_n \tau} G(\vec{q};\tau) , \qquad (5)$$

where $\beta = 1/k_B T$ and $\omega_n = (2n+1)\pi / \beta\hbar$, $n$ being an integer (positive, negative, or zero); for boson-like quantities, such as the polarizability introduced below, $\omega_n = 2n\pi / \beta\hbar$.

In Eq. (4) $G_0(\vec{q};i\omega_n)$ denotes the Green's function for non-interacting electrons; the same is given by the equation

$$G_0(\vec{q};i\omega_n) = \frac{1}{i\omega_n - \omega_{\vec{q}}} , \qquad (6)$$

where the "band energies" are given by the equation

$$\hbar\omega_{\vec{q}} = \hbar^2 q^2 / 2m - \mu , \qquad (7)$$

$\mu$ being the chemical potential. In Eq. (4) we have also introduced the self-energy $\Sigma_{xc}(\vec{q};i\omega_n)$, which can be thought of as (the Fourier transform of) the non-local, time-dependent "potential" in which a "real" (i.e., interacting) electron propagates. We emphasize that the notation employed in Eqs. (4) and (6) is appropriate for a spatially-homogeneous medium (jellium), in which all physical quantities are scalars. Our notation also reflects the fact that, for the homogeneous electron gas, the "tadpole" diagram [43] vanishes (this is why one-electron potentials —e.g., the Hartree potential, or the Kohn-Sham one-electron potential [9]— are not included in either the $G_0$ given by Eq. (6) or in $\Sigma_{xc}$, which here is entirely due to exchange and correlation. For electrons in a periodic crystal, the Dyson equation, Eq. (6), and Eqs. (8) and (9) below, are replaced by matrix equations in either configuration space or in an appropriate basis spanning the Hilbert space of the system; also the tadpole must be accounted for. This more general case is discussed in a forthcoming publication [44].

A key point about Eq. (4) is that the self-energy is a functional of the Green's function $G$ [33,34]. Physically, this means that the particle whose propagation we follow contributes self-consistently to the dynamical "potential" in which it moves. Mathematically, this means that Eq. (4) must be solved self-consistently, i.e., in a "loop" which starts out from the computation of $\Sigma_{xc}$ in terms of $G_0$, followed by a recomputation of the self-energy from an updated Green's function $G$ obtained from Eq. (4), and so on, until convergence is achieved to a desired accuracy. As noted in the introduction, Eq. (4) has been solved self-consistently in recent years for Hubbard-like models with short range interactions [4,5]. This self-consistency has traditionally been ignored for systems with long-range interactions. (A partial degree of self-consistency for electrons in jellium has been reported very recently [45,46]; in the case of semiconductors, self-consistency in the search for the real part of the quasiparticle energies has also been implemented approximately [26,27]).)

We consider the simplest non-trivial self-energy functional which includes the effects of dynamical screening, namely the screened-interaction approximation (SIA) of Baym and Kadanoff [33], in which

$$\Sigma_{xc}(\vec{q};i\omega_m) = -\frac{1}{\hbar\Omega_N}\sum_{\vec{k}}\frac{1}{\beta\hbar}\sum_{i\omega_n}G(\vec{q}-\vec{k};i\omega_m-i\omega_n)V_S(\vec{k};i\omega_n) , \qquad (8)$$

where the screened Coulomb interaction $V_S$ is given by

$$V_S(\vec{q};i\omega_n) = \frac{v(\vec{q})}{1-v(\vec{q})P(\vec{q};i\omega_n)} , \qquad (9)$$

in terms of the dynamical polarizability $P$, which in the present approximation is of RPA form,

$$P(\vec{q};\tau) = 2\sum_{\vec{k}}G(\vec{q}+\vec{k};\tau)G(\vec{k};-\tau) . \qquad (10)$$

Equations (4), (6), (8)–(10) define a self-consistent problem, whose solution yields (implicitly) the self-energy as a functional of the Green's function, $\Sigma_{xc}[G]$, within the SIA. It should be recognized that *the diagram* for the SIA given by Eq. (8) corresponds to the *GW* or *screened-Hartree-Fock* approximation for the self-energy [47,48].

In Eq. (10) we have written down the polarizability for imaginary times, as this is the representation in which we actually calculate this function first. This may seem to be a roundabout way to proceed, in view of the fact that from the Dyson equation we obtain an update for $G(\vec{q};i\omega_n)$, not $G(\vec{q};\tau)$, and, furthermore, it is the Fourier transform of the polarizability, $P(\vec{q};i\omega_n)$, which is required in the screened interaction given by Eq. (9). However, the direct evaluation of $P(\vec{q};i\omega_n)$ as a frequency convolution of two *G*'s (obtained by Fourier-transforming Eq. (10)) converges quite poorly as function of the frequency cutoff which must necessarily be imposed [49]. Thus, we actually use Eq. (10) as it stands. Formally, we generate the Green's function for imaginary times via the inverse of Eq. (5), i.e.,

$$G(\vec{q};\tau) = \frac{1}{\beta\hbar} \sum_{i\omega_m} e^{-i\omega_m \tau} G(\vec{q};i\omega_m) \ . \tag{11}$$

Unfortunately, a direct numerical evaluation of the frequency sum in Eq. (11) is not feasible [49]—this problem is traced to the slow $1/\omega$ decay of the Green's function for large frequencies [50]. We have solved this difficulty by implementing the following physically-motivated procedure. We add and subtract from the argument of the sum in Eq. (11) the Fock Green's function $G_x$, i.e., the Green's function which corresponds to a bare-exchange treatment of the electron-electron interactions; $G_x$ is of the form of Eq. (6), except that the single-particle eigenvalues are shifted by a *frequency-independent* exchange self-energy $\Sigma_x(\vec{q})$. Since the energy-position of the pole of $G_x$ is known, we can evaluate its Fourier transform (Eq. (11)) in closed form. We then have that

$$G(\vec{q};\tau) = \frac{1}{\beta\hbar} \sum_{i\omega_m} e^{-i\omega_m \tau} \{G(\vec{q};i\omega_m) - G_x(\vec{q};i\omega_m)\}$$

$$- e^{-(\omega_{\vec{q}}+\Sigma_x(\vec{q}))\tau} \left(1 - \frac{1}{e^{\beta\hbar(\omega_{\vec{q}}+\Sigma_x(\vec{q}))}+1}\right) \ , \tag{12}$$

where the second term corresponds to the contribution from $G_x$. Of course, the sum in Eq. (12) is evaluated with a finite cutoff; however, since $G_x \approx G$ for large frequencies —physically, this is the case because correlation becomes inoperative at high excitation energies— this sum converges rapidly [49]. Equation (12) is central to our numerical algorithms. From the knowledge of $G(\vec{q};\tau)$ we obtain $P(\vec{q};\tau)$ according to Eq. (10), and the required Fourier coefficients $P(\vec{q};i\omega_n)$ are subsequently obtained via Fast-Fourier-transform techniques.

Finally, we note that the chemical potential $\mu$ (which was introduced into the above scheme through Eqs. (6) and (7)) is not known a priori, since its value is affected by the electron-electron interactions (and the effect is not negligible, as our numerical results show). The renormalization of the chemical potential is self-consistently determined through the implicit equation

$$n = \frac{2}{\Omega_N} \sum_{\vec{k}} \frac{1}{\beta\hbar} \sum_{i\omega_m} e^{i\omega_m \delta} G(\vec{k};i\omega_m|\mu) , \qquad (13)$$

where $n$ is the electron density. Equation (13) is solved for each iteration of Eq. (4). The notation used in Eq. (13) is meant to emphasize the fact that the updated Green's function $G$ depends on the current (updated) value of $\mu$. Of course, as was the case with Eq. (11), we cannot solve Eq. (13) as it stands. Rather, we manipulate the right-hand-side of Eq. (13) in the same way as done above for Eqs. (11) and (12) —this yields a rapidly converging sum over the Matsubara frequencies. Further details will be presented elsewhere [49]

It is instructive to note that, even at just the level of the evaluation of the polarizability, the present approach contains significant advantages relative to the conventional representation given by Eq. (3), which was the basis of the recent progress summarized earlier in this paper [51]. First, because of the pole structure of Eq. (3), an accurate sampling of the Brillouin zone typically requires substantially denser meshes than are required with the present method [49]; this is a major practical advantage, which should become even more relevant in the treatment of more elaborate self-energy functionals.  Second, there is no direct generalization of Eq. (3) once the propagators are dressed by the self-energy (which is the case for all iterations of Eq. (4) beyond the zeroth order, $G = G_0$).

We turn next to a presentation of selected results obtained by the self-consistent implementation of the above scheme.  We confine our discussion to the homogeneous electron gas, and consider the case $r_S = 5$; for this low-density "metal" the effects of correlation are more easily visualized.  All the results presented below correspond to a temperature $T = 800 K$.  The restriction to the jellium model has the conceptual advantage that it allows us to isolate system-independent, or universal, features of the solution of Eq. (4), which, to a greater or lesser extent, should be relevant for all metals.

In Figs. 3 and 4 we show characteristic results for the self-energy, plotted as function of frequency, for a wave vector which lies very close to the Fermi surface.  The fact that this wave vector does not equal $k_F$ *exactly* may seem surprising.  However, as a preamble to the implementation of our methods for real crystals [44], we have actually solved Eq. (4) over a discrete mesh of wave vectors, which are required to satisfy Born–von Karman periodic boundary conditions.  Thus, in essence, we have performed "empty lattice" calculations. (In order to facilitate the comparison with the case of potassium, discussed elsewhere [44], whose bulk density is comparable to the one used in the present calculations, we have assumed a bcc Bravais lattice.) The wave vector $q = 0.99 k_F$ happens to be the closest one to the Fermi surface in our numerical mesh.

Now Figs. (3) and (4) refer to the initial evaluation of the self-energy in terms of the non-interacting Green's function $G_0$, i.e., $\Sigma_{xc}[G_0]$. This corresponds to the level of the calculation of the self-energy of the homogeneous electron gas reported over the years by many authors, starting from the pioneering work of Quinn and Ferrell [47], Hedin [48], and Lundqvist [52]. For the most part, such work has been carried out on the basis of the ground-state $(T = 0K)$ formalism, which yields all physical variables directly on the real-frequency axis. In our case, the data displayed in Fig. 4 were obtained from the data of Fig. 3 via Padé approximants [35]. Basically, the self-energy evaluated over the Matsubara frequencies is fitted to a ratio of polynomials (whose degree is typically of the order of 30, in the calculations reported herein); this allows us to "stretch" the domain of definition of the self-energy elsewhere in the complex frequency plane. In particular, we can perform the analytic continuation to points just above the real axis (10 meV above the real axis, in the present calculations), and obtain the retarded self-energy shown in Fig. 4, which agrees well with previous work [48,52,53]. (A detailed discussion of the impact of various numerical parameters on the analytic continuation will be presented in Ref. [49].)

Figures 3 and 4 are meant to illustrate a point which provides strong motivation for the method we have implemented: While the self-energy is a smooth function of frequency on the imaginary axis, its analytic continuation has sharp structures on the real axis —due, in this case, to resonant coupling to the plasmon [52]. The same qualitative remark applies to all other dynamical quantities, including the polarizability. It is then rather self-suggestive that, as noted above for the particular case of the polarizability, our finite temperature scheme lends itself to a more efficient sampling of the Brillouin zone than the ground-state methods (relatively coarse meshes yield accurate results [49]). Of course, in applications to simple metals, the use of a finite temperature method is not essential; rather, as just noted, we find it to be very practical. On the other hand, in the case of, e.g., magnetic response of transition metals, the temperature plays an essential physical role, which our method is designed to incorporate.

An important physical quantity related to the one-particle Green's function is the spectral function $A(\vec{q};\omega)$, which gives the probability that an added particle with momentum $\hbar\vec{q}$ will find an eigenstate of the interacting $(N+1)$-particle system with energy $\hbar\omega$ [24]. The one-particle spectral function is defined by the equation

$$A(\vec{q};\omega) = -\frac{1}{\pi} \operatorname{Im} G(\vec{q};\omega) , \qquad (14)$$

which we evaluate from the knowledge of the self-energy. Figure 5 shows the spectral function for the same wave vector considered in Fig. 4. The prominent quasiparticle peak at the Fermi surface (energies are measured from the chemical potential), and the well-known satellites which arise as a feedback of the plasmon resonance onto the one-particle spectrum, are clearly observed.

The main point about Fig. 5 is that it illustrates the impact of self-consistency in the solution of the Dyson equation. The dashed line corresponds to the evaluation of $A(\vec{q};\omega)$ from the knowledge of the self-energy obtained in terms of the non-interacting Green's function, i.e., $\Sigma_{xc}[G_0]$; this non-self-consistent result agrees well with previous calculations [48,52,53]. The dashed-dotted line is the spectral function obtained after three iterations through Eq. (4). It is apparent that the spectral weight of the satellites has been substantially reduced. Since the integrated spectral weight is controlled by the sum rule

$$\int_{-\infty}^{+\infty} d\omega \, A(\vec{q};\omega) = 1 \, , \tag{15}$$

whose fulfillment is essential for the probability interpretation of the spectral function, the weight lost by the satellites must go to the quasiparticle —it does. Our numerical solution of Eq. (4) (its analytic continuation) fulfills Eq. (15) to within 0.1%. We note, in passing, that the first-frequency moment sum rule [52] is also fulfilled by our results (in this case, to within 1%); this clearly serves as a powerful check of the overall quality of our numerical solution of the Dyson equation. Finally, the solid line in Fig. 5 shows the spectral function for the converged, self-consistent, Green's function (which, on the scale of the figure, corresponds basically to the sixth iteration of the solution of Eq. (4)). Clearly, the trend noted above for the intermediate (third iteration) step holds all the way to convergence.

We can quantify the reduction in the weight of the satellite structure brought about by the self-consistency by noting the related increase in the weight of the quasiparticle state at the Fermi surface, $Z_k$ [24]. The latter is given by the area under the central peak in Fig. 5; $Z_k = 1$ corresponds to a strict delta-function peak, which is only realized for the non-interacting system. For the three iteration steps considered in Fig. 5 we have, respectively, $Z_k = 0.60$ — which reproduces the "canonical" value reported by Hedin [48] for $r_S = 5$—, $Z_k = 0.73$, and $Z_k = 0.74$. Clearly, the effect of self-consistency is to make the quasiparticles more Sommerfeld-like.

It is also of interest to consider the density of one-particle states, defined by the equation

$$\rho(\omega) = \frac{2}{\Omega_N} \sum_{\vec{k}} A(\vec{k};\omega) . \qquad (16)$$

Now, because of the presence of sharp structures in $A(\vec{k};\omega)$ (while Fig. 5 displays the spectral function as a function of frequency, $A(\vec{k};\omega)$ also contains considerable structure as function of wave vector), the sum required in Eq. (16) must be performed over a dense $\vec{k}$-mesh —much denser than the one for which we need to solve Eq. (4) to obtain converged results for, e.g., the quasiparticle weight, or the chemical potential. This computational requirement was handled by performing a cubic-spline interpolation of the (real-$\omega$) self-energy from the coarser $\vec{k}$-mesh for which we solve the self-consistency problem to the denser $\vec{k}$-mesh required by Eq. (16). (We would like to mention, in passing, that this procedure required an "automated" use of Padé approximants for the entire mesh for which Eq. (4) was solved; fortunately, this caused no problems, despite the notorious "instabilities" associated with Padé methods.)

The calculated density of states is shown in Fig. 6. The dashed-dotted line is the $\rho(\omega)$ which corresponds to the initial computation of the self-energy, $\Sigma_{xc}[G_0]$; this curve agrees well with Lundqvist's original *GW* results [52] (the slight rounding of features in the satellite structure compared to the corresponding curve in Ref. [52] is due to the analytic continuation via Padé). The solid line represents the converged result, obtained from the self-consistent result for spectral function shown in Fig. 5. For reference, in Fig. 6 we also show the density of states for non-interacting electrons. The large reduction of the weight of the satellite structure in $\rho(\omega)$ observed in Fig. 6 is consistent with the results for the spectral function, and the quasiparticle weight, discussed above.

A related feature of the calculated $\rho(\omega)$ needs to be addressed. This is the widening of the occupied bandwidth observed in Fig. 6 (note that the zero of energy is the renormalized chemical potential for the corresponding solution of Eq. (4)). This result, which is intimately connected with the reduction in the weight of the satellites brought about by self-consistency, seems to be contrary to observation [54]; note, however, that there is no clear experimental evidence for the existence of intrinsic many-body satellites in simple metals. A related issue is the fact that, for small wave vectors, the renormalized polarizability $P$ differs qualitatively from the bare polarizability $P^{(0)}$ for frequencies in the plasmon region; this question will be discussed in more detail elsewhere [49]. It is, of course, possible that the inclusion of vertex corrections —which at least to low order in perturbation theory have been shown to "counteract" the effect of the self-energy insertions [55]— will alter the above results, yielding, for example, a reduction in the occupied bandwidth, and a concomitant increase in the weight of the satellites. We will explore more general self-energy functionals in future work. Nonetheless, our present results indicate that a self-consistent treatment of the interactions is required in order to obtain quantitatively reliable results.

## ACKNOWLEDGMENTS

W.-D. S. gratefully acknowledges a postdoctoral fellowship from the Deutsche Forschungsgemeinschaft. This work was supported by National Science Foundation Grant No. DMR-9634502 and the National Energy Research Supercomputer Center. Oak Ridge National

FIGURE CAPTIONS

Fig. 1.   Measured many-body local field factor, $G(\vec{q},\omega)$, for Al [23]. The full (empty) symbols correspond to the real (imaginary) part of $G$, respectively. The local-field factors calculated (for jellium) in Refs. [38] (UI), [39] (RA), [40] (BDL), and [41] (MCS) are also shown, as is the corresponding LDA result.

Fig. 2.   Plasmon dispersion relation in Cs for small wave vectors along the (110) direction [14]. The 5$p$ semi-core states were treated as valence states. Right (left) panel includes (ignores) the crystal local fields. Theoretical curves are labeled by the vertex correction used in the respective calculation —see text.

Fig. 3.   Electron self-energy $\Sigma_{xc}(\vec{q};i\omega_n)$ for $q = 0.99 k_F$, plotted as function of the continuous variable $i\omega$. The figure refers to a first evaluation of the self-energy, i.e., $\Sigma_{xc} = \Sigma_{xc}[G_0]$, for $r_S = 5$.

Fig. 4.   Analytic continuation of the electron self-energy of Fig. 3 to real frequencies. It is instructive to compare the sharp spectral features observed in this figure with the smooth frequency dependence shown in Fig. 3.

Fig. 5.   Spectral function for one-particle excitations, $A(\vec{q};\omega)$, for $q = 0.99 k_F$. The figure shows the spectral function obtained from a first evaluation of the self-energy (dashed line), from the third iteration of the solution of Eq. (4) (dashed-dotted line), and from the self-consistent solution of Eq. (4) (solid line).

Fig. 6.   Density of one-particle states, $\rho(\omega)$, for $r_S = 5$. The figure show the $\rho(\omega)$ which corresponds to a first evaluation of the self-energy, i.e., $\Sigma_{xc} = \Sigma_{xc}[G_0]$, and to the converged, self-consistent solution of Eq. (4). For reference, the $\rho(\omega)$ for non-interacting electrons is also shown.

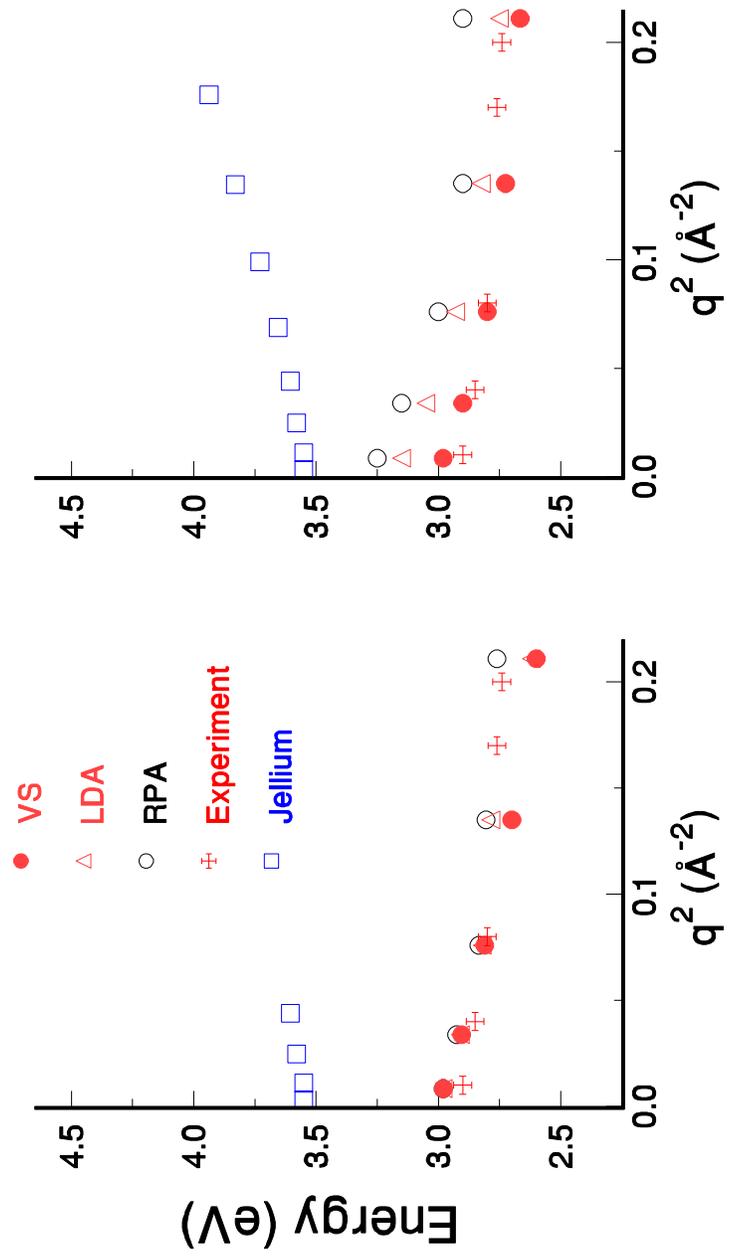

Fig. 2

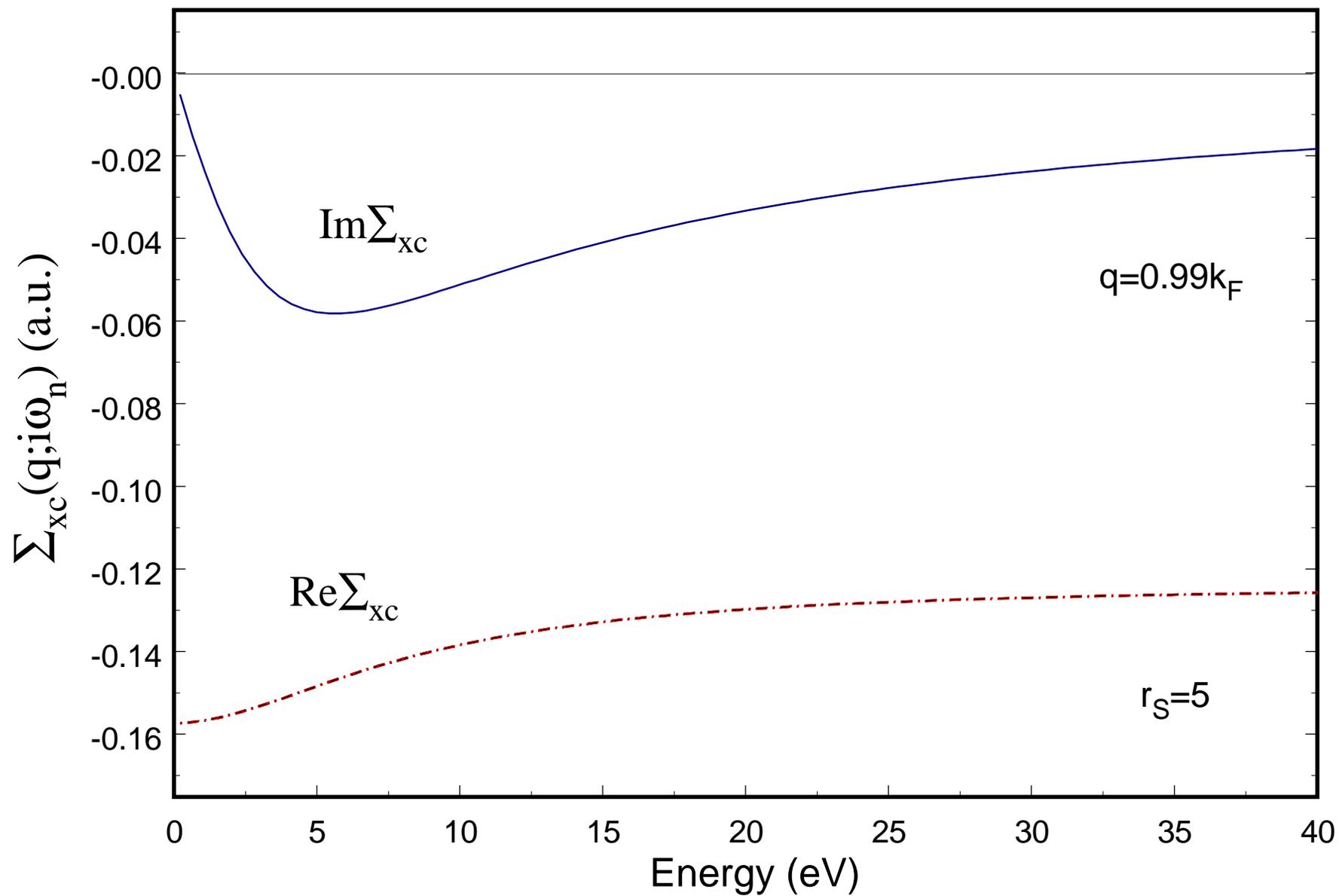

Fig. 3

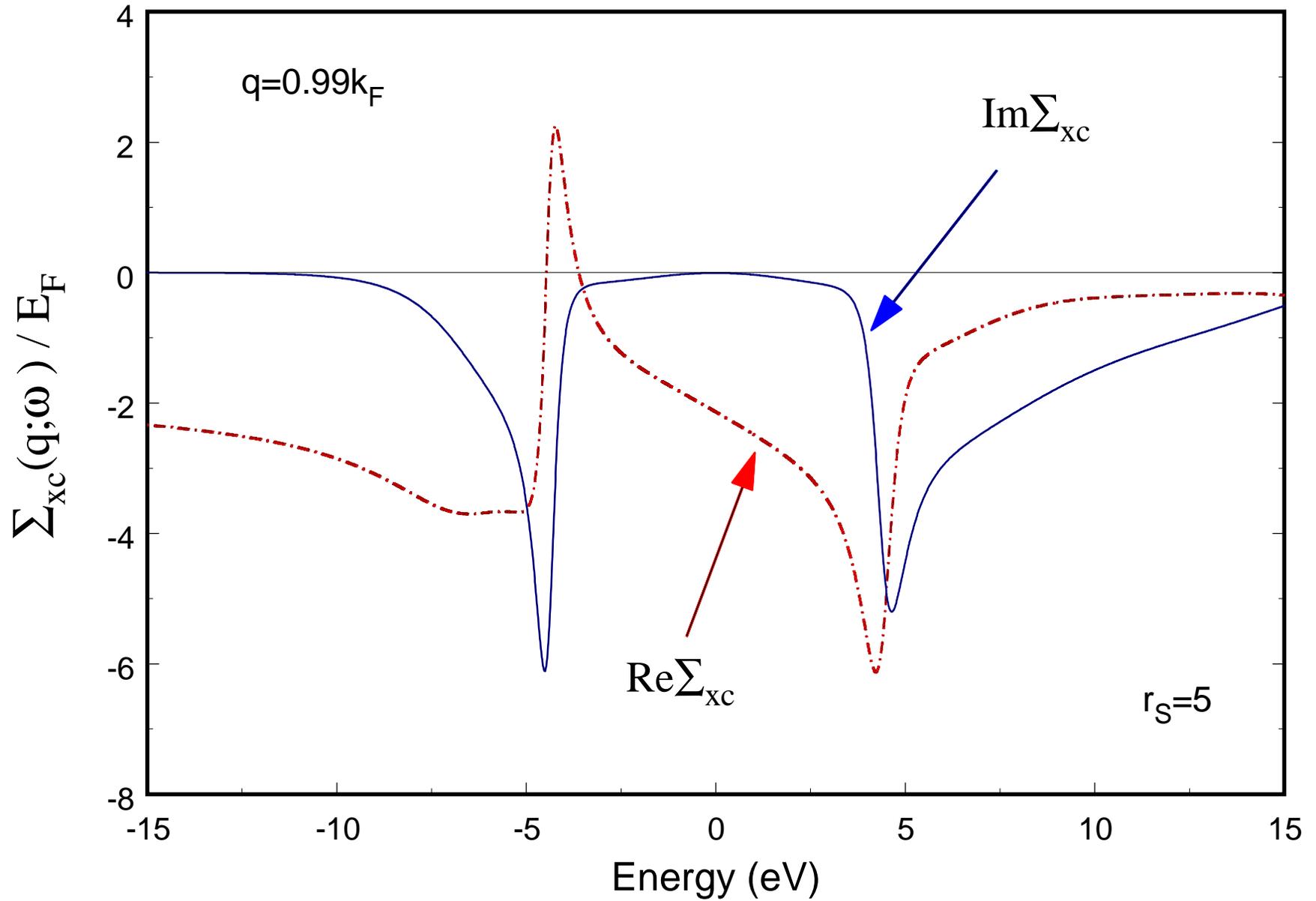

Fig. 4

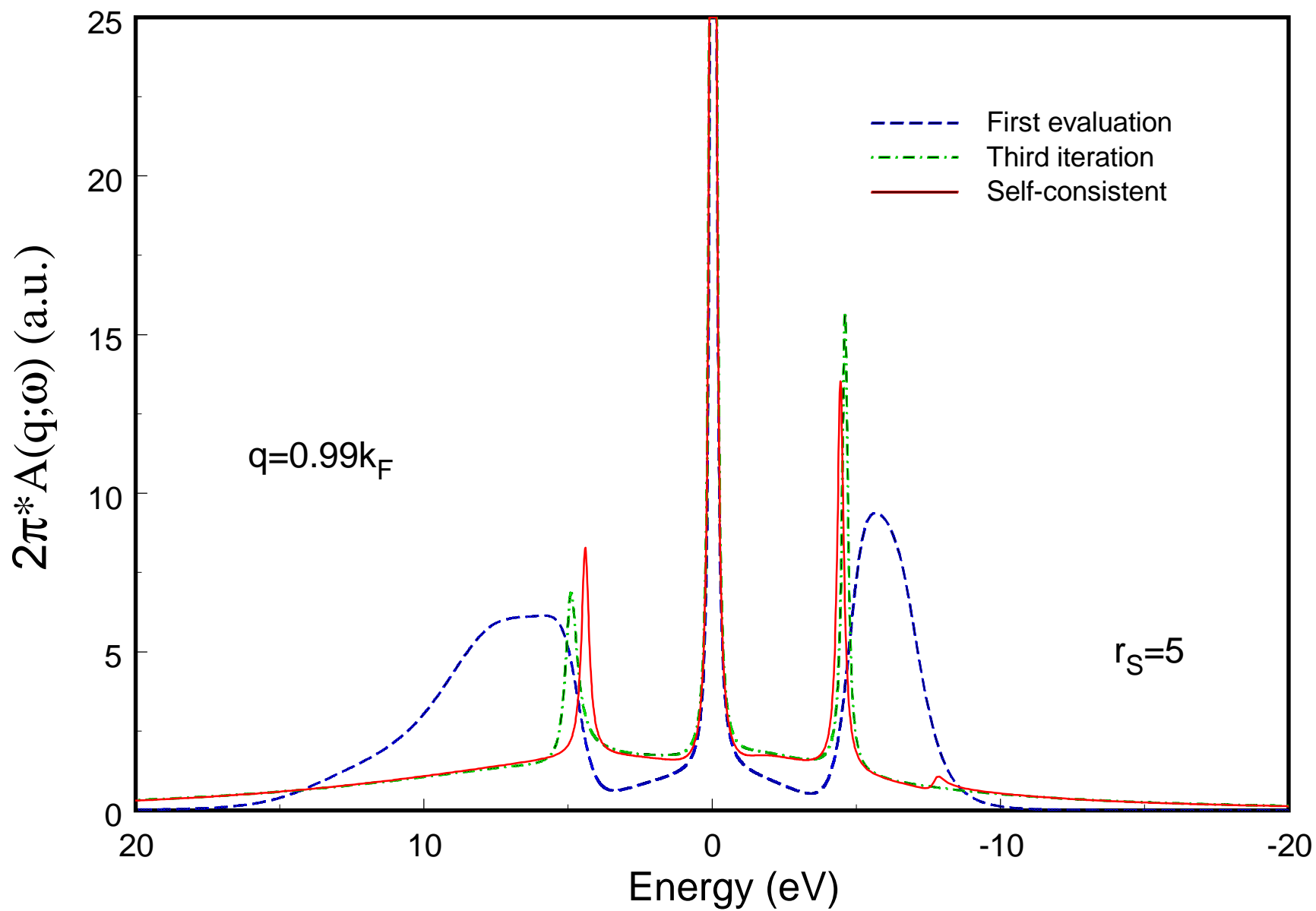

Fig. 5

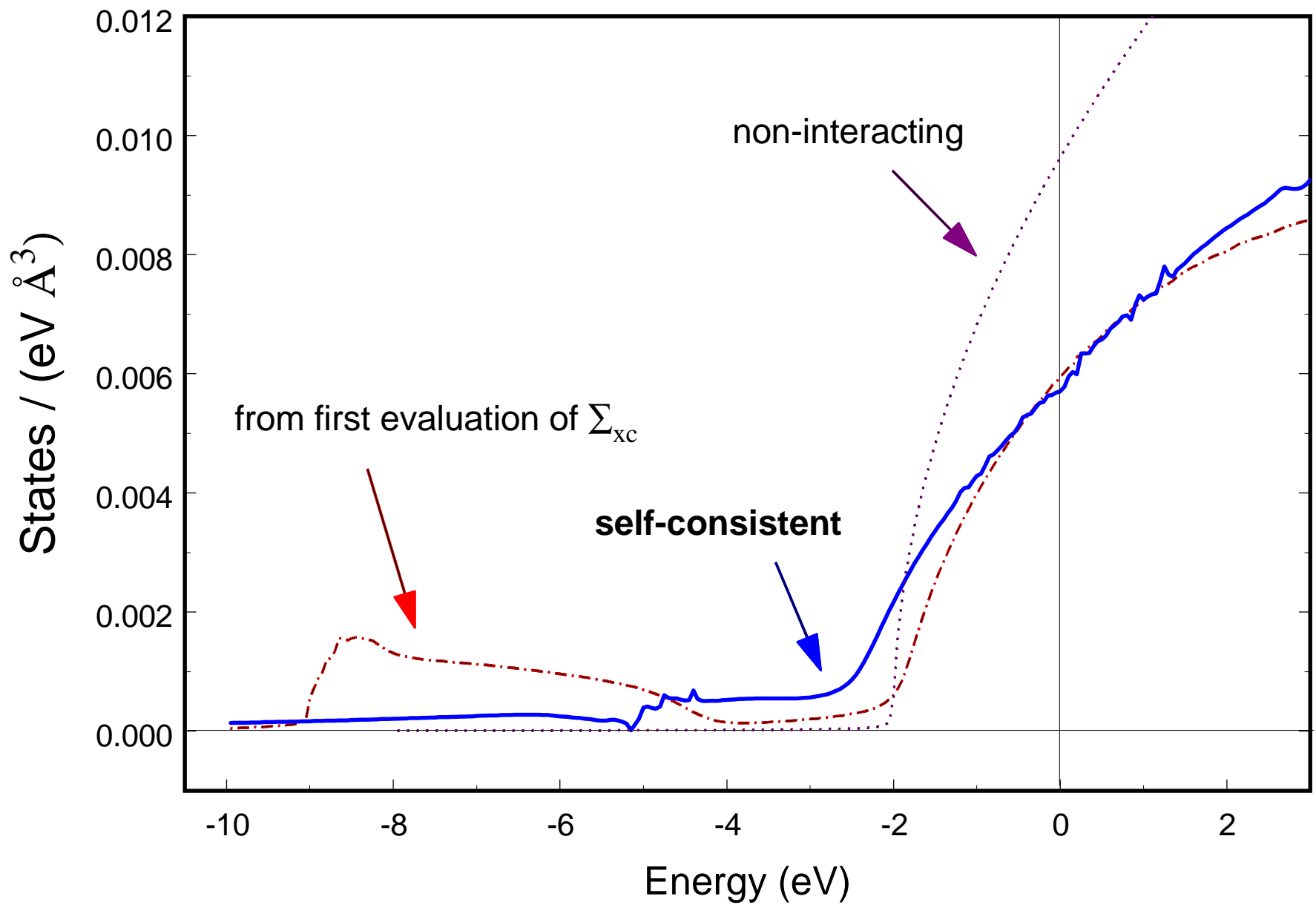

Fig. 6